\documentclass[preprint,prd,groupedaddress,footinbib,showpacs,rotate]{revtex4}

\usepackage{graphicx}
\usepackage{dcolumn}
\begin{document}
%\hfill hep-th/yymmnnn   
   
\hfill MCTP-07-17\\

\title{ On a Holographic Model for Confinement/Deconfinement}
\author{C.  A.  Ballon Bayona}
\email{ballon@if.ufrj.br} \affiliation{Instituto de F\'{\i}sica,
Universidade Federal do Rio de Janeiro, Caixa Postal 68528, RJ
21941-972 -- Brazil}
\author{Henrique Boschi-Filho }
\email{boschi@if.ufrj.br} \affiliation{Instituto de F\'{\i}sica,
Universidade Federal do Rio de Janeiro, Caixa Postal 68528, RJ
21941-972 -- Brazil}
\author{Nelson R. F. Braga}
\email{braga@if.ufrj.br} \affiliation{Instituto de F\'{\i}sica,
Universidade Federal do Rio de Janeiro, Caixa Postal 68528, RJ
21941-972 -- Brazil}
\author{Leopoldo A. Pando Zayas}
\email{lpandoz@umich.edu} \affiliation{ Michigan Center for Theoretical Physics, 
Randall Laboratory of Physics, The University of Michigan,  
Ann Arbor, MI 48109-1120}

%\date{\today}

\begin{abstract} 
We study the thermodynamics of the hard wall model, which consists in the introduction of an infrared cut-off
in asymptotically $AdS$ spaces. This is a toy model for confining backgrounds in the 
context of the gauge/gravity correspondence. We use holographic renormalization and reproduce the 
existence of a Hawking Page phase transition recently discussed by Herzog. We also show that the entropy 
jumps from $N^0$ to $N^2$, which reinforces the interpretation of this transition as the gravity dual 
of confinement/deconfinement. We also show that similar results hold for the phenomenologically motivated 
soft wall model, underlining the potential universality of our analysis. 

\end{abstract}


\maketitle

\section{ Introduction }

A remarkable example of a gauge/string duality is the Maldacena conjecture, known as 
AdS/CFT correspondence\cite{Maldacena:1997re,Gubser:1998bc,Witten:1998qj}, that proposes an equivalence between 
string theory in some ten dimensional curved spaces and superconformal gauge theories in lower dimensional spaces. In particular, string theory in $AdS_5 \times S^5 \,$ space is dual to four dimensional 
 ${\cal N}\,=4\,$ supersymmetric $SU(N) $ Yang Mills theory. This correspondence has been extended to 
more realistic situations and has been shown to provide a supergravity description of field theories that display confinement 
and chiral symmetry breaking \cite{Klebanov:2000hb,Maldacena:2000yy}. 

At finite temperature, there are many interesting phenomena that can also be studied using this correspondence. 
Witten \cite{Witten:1998zw}  used the AdS/CFT correspondence 
to relate the thermodynamics of  gauge theories to the thermodynamics of asymptotically anti-de Sitter ($AdS$) spaces. The main result is 
an interpretation of the work of Hawking and Page \cite{Hawking:1982dh}. 
Namely, Hawking and Page studied the thermodynamics of four dimensional asymptotically  $AdS$ spaces  
with compact boundary $S^2 \times S^1$ . They considered two different solutions of the Einstein equations:
the $AdS$ and the black hole $AdS$ spaces. In a quantum gravity theory the path integral should involve 
contributions from all spaces with a fixed asymptotic boundary. In a semiclassical approximation, the 
dominant contribution comes from the classical solution of minimum action. Using this  approach, Hawking 
and Page found a phase transition at some critical temperature.
Above this temperature the black hole solution is thermodynamically preferred. For lower temperatures 
the (thermal)  $AdS$ solution is dominant. 

 Witten extended this analysis to the cases 
of $n+1$ dimensional spaces with  $S^{n-1} \times S^1$ (compact) and $R^{n-1}  \times S^1$ (non-compact) 
boundaries. 
For the $S^{n-1} \times S^1$ case the gravitational solutions are those of Hawking and Page
and there is a thermal phase transition. 
From the point of view of the dual gauge theory this is argued to be a deconfinement phase transition.
For the non-compact spatial boundary $\,R^{n-1} \times S^1\,$  the classical solutions are  $AdS$ and  black hole $AdS$ in Poincar\'e coordinates. In this case the black hole $AdS$ space is dominant  
for all non zero temperatures and so there is no thermal phase transition. 

Confinement in gauge theories can be understood studying the behavior of Wilson loops.
The AdS/CFT correspondence can be used to calculate the string configurations dual to a Wilson loop.
For a gauge theory dual to  Poincar\'e $AdS$ space (non compact boundary) it has been shown 
that the potential energy of a pair of static charges has a Coulombic behavior \cite{Rey:1998ik,Maldacena:1998im}. 
Witten \cite{Witten:1998zw} suggested that, for the case of compact boundary, the $AdS$ space corresponds to the 
confining phase 
while the black hole in $AdS$ phase corresponds to the deconfined phase.  This lead him to associate the Hawking Page transition with
the deconfinement phase transition.  In the case of non compact boundary both spaces
are non confining. 
The Wilson loops for black hole $AdS$ space with non compact boundary have been explicitly 
calculated in \cite{Rey:1998bq,Brandhuber:1998bs}. 
A general criteria for confinement based on Wilson lines was presented  in \cite{Kinar:1998vq}. 

The criteria for confinement motivates some simple toy models of {\it bona fide} type IIB supergravity backgrounds dual to 
confining gauge theories. Particularly interesting has been the hard wall model which consist of Poincar\'e 
$AdS$ space with a hard infrared cut-off. Some interesting results of QCD have been recently reproduced from 
gauge string dualities using this 
model. For example, Polchinski and Strassler introduced an infrared cut off in Poincar\'e $AdS$ space
to obtain the scaling of hadronic high energy scattering amplitudes at fixed angles from string theory 
\cite{Polchinski:2001tt}. 
The hard wall model was also useful for calculating masses in the hadronic 
spectrum\cite{Boschi-Filho:2002vd,deTeramond:2005su,Erlich:2005qh, DaRold:2005zs}. 

In above examples one might argue that the main role was played by conformal invariance or other symmetries and 
therefore the results bear a kinematic undertone, depending lightly on the cut off. However, it 
was shown recently by Herzog \cite{Herzog:2006ra} that the introduction of 
infrared cut-offs in asymptotically $AdS$ spaces with  non compact 
boundary $R^{3}  \times S^1$  leads to a Hawking-Page thermal phase transition at a finite critical temperature.
Below this temperature the thermodynamically dominant space is $AdS$ and above it is a black hole in $AdS$. 
The dual gauge theory should undergo a deconfinement phase transition.

The aim of this letter is to revisit this transition. We focus on a rigorous approach to the gravity action and on the question of the number of degrees of freedom involved. 
The gravity action involves surface terms besides the standard Einstein Hilbert bulk 
action \cite{Gibbons:1976ue}. 
We will see here that these surface terms do not vanish for  $AdS$ and black hole $AdS$ in Poincar\'e coordinates. Both volume and surface terms are infinite and must be regularized. 
One possible way of defining a finite action is by subtracting a reference background.
A more general procedure is known as the holographic renormalization. It mimics field theory renormalization by 
including boundary counterterms. In the case of interest to us, it has been neatly described in 
\cite{Balasubramanian:1999re,Emparan:1999pm}. We will use this approach here to construct finite actions 
for the $AdS$ and black hole $AdS$ spaces in the hard wall model. As a bonus, we are also able to understand the change in the number of degrees 
of freedom by calculating the corresponding entropies. 

A deconfinement phase transition is associated with a sudden increase in the number of degrees of freedom of the theory. In Yang Mills $SU(N)$ gauge theories in a deconfined phase, where the gluons are free, the number of degrees of freedom is proportional to $N^2\,\,$. In a confined phase, where only color singlet states are allowed, the number of degrees of freedom is proportional to $N^0$. Here we calculate the entropies for the
$AdS$ and black hole $AdS$ spaces in the hard wall model and find the expected jump in the entropy at the phase transition.
 
This article is organized as follows. 
In section {\bf II} we calculate the gravity actions including the surface terms.
We use the method of adding counterterms and find finite actions.
In section {\bf III} we consider the phase transition, with fixed  hard cut-offs in the spaces, and 
calculate the corresponding entropies. In section {\bf IV} we present a brief discussion of the soft 
wall model.
Then we present some final remarks in section {\bf V}.

\section{ Gravity action,  surface terms and counterterms}

The gravity action for five dimensional empty space is

\begin{equation}
I \,=\, - \frac{1}{2\kappa^2}\, \int_{{\cal M}} d^5 x
\sqrt{g}\,\Big( {\cal R} \,-\,\Lambda \Big) - \frac{1}{\kappa^2}\,
\int_{{\cal \partial M}} d^4 x \sqrt{h} K \,\,, \label{Action1}
\end{equation}

\noindent where $\kappa^2\,$ is proportional to the five dimensional Newton constant 
($ \kappa^2 = 8 \pi G_5\,\,$) , $\cal{R}$  is the Ricci scalar,  $\Lambda $ the
cosmological constant,
 $K$ is the trace of the extrinsic curvature of the boundary and $h$ is the
 determinant of the boundary induced metric $h_{\mu \nu}\,$.
The first integral is a bulk term which is the familiar Einstein
Hilbert action that leads to the equation of motion

\begin{equation}
{\cal R}_{ab}-\frac{1}{2}g_{ab}\, {\cal R}  = -\frac{1}{2}\Lambda g_{ab}\, . 
\label{Ricci}
\end{equation}

 The second  is a surface term, introduced by Gibbons and
Hawking \cite{Gibbons:1976ue}, coming from the variational principle. The bulk and surface terms
in the gravity action are typically divergent. An intuitive 
attempt to deal with this problem is to subtract from
(\ref{Action1}) the contribution of a reference background 
spacetime with the same boundary. However, there are many cases
in which this subtraction is not possible, as exemplified by rotating black hole solutions. 
A more rigorous approach has been developed for asymptotically AdS 
spacetimes \cite{Balasubramanian:1999re,Emparan:1999pm}.  The
AdS/CFT correspondence relates the gravity action of these
spacetimes with the quantum effective action of the dual conformal
field theories defined on their boundaries. The divergences on the gravity action are then 
interpreted as UV divergences of the quantum field theory. 
These UV divergences can be removed by
adding local counterterms to the action. The counterterm action
 can be expressed  in the following form\cite{Emparan:1999pm}:

\begin{equation}
I_{ct} \, = \, \frac{1}{\kappa^2}\, \int_{{\cal \partial M}} d^4 x
\sqrt{h} \,  F (R , \tilde{ {\cal R}}, \nabla \tilde{ {\cal R}})
\, ,  \label{gencounter}
\end{equation}

\noindent where $F$ is a finite series of diffeomorphism invariants constructed
from the AdS radius $R$ and the boundary induced curvature
$\tilde{ {\cal R}}$. The terms in $F$ are fixed by requiring
finiteness of the total gravity action: \, $I _{total} \, = \, I
+ I _{ct}$ . This procedure is known as holographic renormalization.

In this work we will study  two asymptotically AdS spaces which
have the non compact boundary $R^3 \times S^1 \, $ :  thermal AdS
and black hole AdS in Poincar\'e coordinates.  These spaces are solutions of the equation of 
motion (\ref{Ricci}), with negative cosmological constant $\Lambda = -
12/R^2\,$. They have a compact time parametrizing \, $S^1\,$,
 \, with a period related to the temperature of the conformal field
theory on the boundary. We will introduce a hard cut-off in these spaces
and look for their thermodynamical properties.

The thermal AdS space is an Euclidean AdS with compact time
coordinate $\tau\,$ with period $\beta\,$ and temperature $ T = 1/
\beta\,$. The metric is

\begin{equation}
\label{TAdS} ds^2 = \frac{R^2}{z^2}( dz^2 +   d\tau^2 +
d\overline{x}^2 ) \, .
\end{equation}

\noindent We introduce a  cut-off $z_0$ \,  in the space
restricting the radial coordinate  $z$ to  $ 0 < z \le z_0\,$.
This corresponds, from the point of view of the boundary dual gauge theory, to an 
infrared cut-off in the energies proportional to $ 1/z_0\,$.

The black hole AdS space with compact Euclidean time $\tau^\prime
$ with period $\beta^{\prime} = 1/ T^{\prime}$ is

\begin{equation}
ds^2 = \frac{R^2}{z^{\prime\,2}} \left(
\frac{dz^{\prime\,2}}{f(z^{\prime})} +
f(z^{\prime})d\tau^{\prime\,2} + d{\overline{x}^{\prime}}^2
\right) \, . \label{AdSBH}
\end{equation}

\noindent where $f(z^{\prime})\equiv
1-\frac{z^{\prime\,4}}{z_{h}^4}$ and $z_h\,$ is the horizon position.
Introducing a hard cut-off $z_0^\prime$ as a maximum value for $z'$
we have two cases:  $z_0^\prime \le z_h$ or $z_0^\prime \ge z_h$.
In the first case we restrict $z'$ to $ 0 < z^\prime \le z_0^\prime\,$.
In the second case there will be no cut-off in this space and $ 0 < z^\prime
\le z_h \,$. So, in general we have $ 0 < z^\prime \le \bar z \,$ with 
$ \bar z \,=\, {\rm min} (z_0^\prime , z_h) \,$.

The condition of non singular behavior of the metric
on the horizon $ z^{\prime} = z_{h}$ leads to a relation between
the horizon position and the temperature: $ 1/T^\prime =
\beta^\prime = \pi z_h $. Note that the periods $\beta $ and $
\beta^\prime $ do not depend on the radial coordinates $z$ or $z^\prime$. 
Then, in order to have the same boundary the periods $\beta $ and $
\beta^\prime $ must be equal. 

The bulk term of the action for the $AdS$ and black hole $AdS$ spaces, 
after substituting the value of the Ricci scalar $
{\cal R} = 5 \Lambda /3 \,= -20/R^2\, $,  takes the form

\begin{equation}
I_{bulk} \,=\,  \frac{4}{ R^2\,\kappa^2}\, \int_{{\cal M}} d^5 x
\sqrt{g} \,. \label{Action2}
\end{equation}

There is a trivial integration over the $x^i$ coordinates. We can
define a new quantity, an action density, that we will still
represent by $I$ dividing  by this common (infinite) factor. Since
there is a singularity in the metrics at $z = z^\prime =  0$ we
introduce small UV regulators $\epsilon$ and $\epsilon^\prime$. The
bulk actions take the form

\begin{eqnarray}
I^{^{AdS}}_{bulk} &=& \frac{4R^3}{\kappa^2}\, \int_0^\beta d\tau
\int_\epsilon^{z_0} dz \,z^{-5} \,=\,
\frac{R^3}{\kappa^2}\,\beta\,\Big( \frac{1}{\epsilon^4 } -
\frac{1}{z_0^4 } \Big) \label{Vol1}
\\
\nonumber\\
I^{^{BH}}_{bulk} &=& \frac{4R^3}{\kappa^2}\,  \,
\int_0^{\beta^\prime} d\tau \int_{\epsilon^\prime}^{ \bar z } dz \,z^{-5}\,=\,
\frac{R^3}{\kappa^2}\,\beta^\prime \, \Big( \frac{1}{{\epsilon'}^4
} - \frac{1}{{\bar z}^4 } \Big) \label{Vol2}
\end{eqnarray}

\noindent where $ \bar z =  {\rm min} (z_0^\prime , z_h) \,$.

Let us now calculate the surface terms of the gravity action

\begin{equation}
I_{surface} \,=\, - \frac{1}{\kappa^2}\, \int_{{\cal \partial M}}
d^4 x \sqrt{h} K \,\,, \label{SurfAction0}
\end{equation}

\noindent for the AdS and black hole AdS spaces. The trace of the
extrinsic curvature at a boundary hypersurface is given by 
\cite{Wald:1984rg,Arutyunov:1998ve}

\begin{equation}
K \,=\, \nabla_a\, n^a \, =\frac{1}{\sqrt{ g}}\,\,\partial_a\,
\Big( \sqrt{ g} \,\,n^a\Big)\,\,,
\end{equation}

\noindent where $n^a$ is a unitary vector normal to the boundary.  

We assume that the regions  $z = z_0$ and $z' = z'_0$ are not part of the boundary of these spaces.
The introduction of cut-offs in these spaces is a model for realizing confinement.
The condition of confinement is that the product of the metric components $ g_{\tau\tau} g_{xx}$ has a minimum that is different from zero \cite{Kinar:1998vq}. 
This behavior shows up naturally in the Klebanov Strassler metric
dual to ${\it{N}}=1$ SYM \cite{Klebanov:2000hb}.
In this space the minimum of the metric appears in a region (corresponding to $z = z_0$ or $z' = z'_0$)
where this metric is not singular. So, this region is not part of the boundary in this more realistic model.
Inspired by this fact,  we will not include the regions $z=z_{0} $ and $z^\prime=z_0^\prime  \,$ as part of  boundaries.

Then the  $AdS$  boundary is the region $z=\epsilon $ and the black hole $AdS$ boundary is
$z^\prime =\epsilon^\prime $. 
For the $AdS$ we have $n^a \,=\, (-z/R, 0, 0, 0, 0)$ and $\sqrt{h}\,=\,R^4 /z^4 \,$. 
For the black hole $AdS$  $n^a \,=\,  ( -z^\prime \sqrt{f(z^\prime)} /R, 0, 0, 0, 0)$ and
$\sqrt{h}\,=\,R^4 \sqrt{f(z^\prime)} / {z^\prime}^{4}\, $.
With these informations we calculate the
surface term (\ref{SurfAction0}). Dividing by the volume factor coming from the $x^i$ coordinates, we find

\begin{eqnarray}
I_{surface}^{AdS} &=& -\, \frac{4 R^3}{\kappa^2}\,
\frac{\beta}{\epsilon^4}  \, , \label{SurfAction1}
\\
I_{surface}^{BH} &=& -\,\frac{4 R^3}{\kappa^2}\,\beta'\, \Big(
\frac{1}{{\epsilon^\prime}^4} \,-\,\frac{1}{2{z_h}^4} \,  \Big)\,
\, . \label{SurfAction2}
\end{eqnarray}

Comparing these actions with the bulk actions  (\ref{Vol1}) and
(\ref{Vol2}) we see that the divergent part of the surface terms
are just $ (- 4) $ times the divergent part of the bulk terms.
This is consistent with the general analysis of ref.
\cite{Emparan:1999pm}. So, the gravity action has a divergent term
 of  $ - 3 R^3 \beta /\kappa^2 \epsilon^4$ \, for AdS and
 $ - 3 R^3 \beta^\prime /\kappa^2 {\epsilon^\prime}^4$ for black hole AdS. 
These divergencies must be canceled by the counterterm actions given by eq. 
(\ref{gencounter}). This determines uniquely the function $F(R) = 3/R$. 
Then, we have 

\begin{equation}
I_{ct} \, = \, \frac{1}{\kappa^2}\, \int_{{\cal \partial M}} d^4 x
\sqrt{h} \, \frac{3}{R} 
\, . \label{gencounter2}
\end{equation}

So, dividing by the infinite volume associated with the coordinates $x^i$  we find 

 \begin{eqnarray}
I_{ct}^{AdS} & = & 3
\frac{R^3}{\kappa^2}\frac{\beta}{\epsilon^4} \, , \label{counter1}
\\
I_{ct}^{BH} & = & 3
\frac{R^3}{\kappa^2}\frac{\beta^\prime}{{\epsilon^\prime}^4}\sqrt{f(\epsilon^\prime)}\,
\simeq \,  3 \frac{R^3}{\kappa^2} \, \beta^\prime\, \Big
(\frac{1}{{\epsilon^\prime}^4} - \frac{1}{2z_{h}^4} \Big ) \, .
\label{counter2}
 \end{eqnarray}

With all these results, the total actions for thermal AdS and black hole AdS are

\begin{eqnarray}
I_{total}^{AdS} &=&  - \frac{R^3}{\kappa^2} \,
\frac{\beta}{z_{0}^4} \, ,
\\
I_{total}^{BH} &=&  - \frac{R^3}{\kappa^2} \, \beta^\prime \,
\Big ( \frac{1}{{\bar z}^4} - \frac{1}{2z_{h}^4} \Big ) \,  ,
\end{eqnarray}

\noindent where again $ \bar z =  {\rm min} (z_0^\prime , z_h) \,$.  These actions are finite as expected. 

\section{ Confinement/deconfinement phase transition}

We now calculate the difference of the regularized actions. 
Setting  $\beta =
\beta^\prime$ \, ,we find
\begin{equation}
I_{total}^{BH} - I_{total}^{AdS} \, = \, -
\frac{R^3}{\kappa^2} \, \beta \Big ( \frac{1}{{\bar z}^4} -
\frac{1}{2z_{h}^4} - \frac{1}{z_{0}^4} \Big ) \, .
\end{equation}

We can separate the two cases $z_0^\prime < z_{h}$ and
$z_0^\prime > z_{h}$:

\begin{equation}
\label{totaldiff} I^{^{BH}}_{total}\,-\, I^{^{AdS}}_{total}\,=
\left\{
\begin{array}{c}
 -
\frac{R^3}{\kappa^2} \, \beta \Big ( \frac{1}{{z_0^\prime}^4} -
\frac{1}{2z_{h}^4} - \frac{1}{z_{0}^4} \Big )
\,\,\,\,(z_0^\prime < z_h)\\
 -
\frac{R^3}{\kappa^2} \, \beta \Big (  \frac{1}{2z_{h}^4} -
\frac{1}{z_{0}^4} \Big )
 \,\,\,\,(z_0^\prime > z_h )\,\,.
\end{array}
\right.
\end{equation}

If we follow the approach of \cite{Herzog:2006ra} and set
 $z_0^\prime = z_0$ we get the same result

\begin{equation}
\label{totaldiffherz} I^{^{BH}}_{total}\,-\, I^{^{AdS}}_{total}\,=
\left\{
\begin{array}{c}

\frac{R^3}{\kappa^2} \, \beta \Big ( \frac{1}{2z_{h}^4} 
 \Big )
\,\,\,\,(z_0^\prime < z_h)\\
 -
\frac{R^3}{\kappa^2} \, \beta \Big (  \frac{1}{2z_{h}^4} -
\frac{1}{z_{0}^4} \Big )
 \,\,\,\,(z_0^\prime > z_h )\,\,.
\end{array}
\right.
\end{equation}

So there is a transition at $ z_0^4\,= \,2 z_h^4\,$, corresponding to 
a critical temperature $T_C =  2^{1/4} /\pi z_0 \,$.
For lower temperatures the  $AdS$ action is smaller, so this space 
dominates the saddle point approximation. For higher temperatures the black hole space dominates.

Now we calculate the thermodynamical variables for these spaces within this semiclassical approximation.
The free energy is  $\, F \,=\, - T \,log Z \,\simeq \, T \, I_{total}\,\,$.
So, using $\kappa^2 = 8 \pi G_5 $  we find 

\begin{eqnarray}
F_{AdS} & = &  - \frac{R^3}{ 8 \pi G_5 \,z_{0}^4} \,\,\,\,\,\,\,\,\,\,\,\,(T < T_C )\, ,
\\
F_{BH} & = &   
 - \frac{R^3}{ 16 G_5 } \,  \pi^3 T^4    \,\,\,\,\,\,\,\,\,\,\,\, (T > T_C)\,\,.  
\end{eqnarray}

Let us investigate the dependence on $N$ of the free energies to compare with the $ SU(N) $ 
Yang Mills theory. For the black hole space, the five dimensional Newton constant $G_5$  
is related to the fundamental string scale by:

\begin{equation}
\label{G5}
G_5 R^5 \,= \, 8 \pi^3 g^2 \alpha'^4\,,
\end{equation}

\noindent where $g $ is the string coupling ($= g_{YM}^2\,)$. The $AdS$ radius satisfies $R^4 \,=\, 4 \pi g N\,{\alpha'}^2$. So we can express the result for the free energy at high temperatures as

\begin{equation}
F_{BH} \, = \,-\, \frac{ N^2 \pi^2}{8} \,T^4  \,\,.
\end{equation}

\noindent This result  agrees with \cite{Gubser:1996de,Gubser:1998nz} and corresponds 
to the free energy of the ${\cal N} = 4\,$ super Yang Mills at strong coupling.

For the $AdS$ with the infrared cut-off  we can not use relation (\ref{G5}). 
We have to draw inspiration from the infrared behavior of {\it bona fide} confining backgrounds. 
For example, in the infrared region of the Klebanov Strassler space\cite{Klebanov:2000hb} 
we can perform a dimensional reduction on the five-dimensional angular part of the 10-d metric. 
Since we have a finite three-sphere at the end of the conifold we expect that the five dimensional volume would be proportional to a third power of the deformation parameter and two powers of the corresponding $AdS$ radius. Namely, $V_5\sim z_0^3\,R^2$ up to some numerical factors. Inspired by this relation we assume that 
our 5-d Newton constant is related to the string scale by
\begin{equation}
G_5 z_0^3 R^2 \sim g^2 \alpha'^4\,.
\end{equation}

\noindent So, the free energy takes the form
\begin{equation}
\label{FADS}
F_{AdS} \, \sim \,-\, \frac{ N^2 }{z_0 R^3}  \,\,.
\end{equation}

In the 't Hooft limit $g N$ is fixed for $N \to \infty$. So the AdS radius 
$R \,=\,( 4 \pi g N\,{\alpha'}^2)^{1/4} $ is independent of $N$. It remains 
to determine the $N$  dependence of $z_0$. This is possible by looking at the effective string 
tension $\sigma \,$.  Consider a quark anti-quark pair separated by a distance
$L$. For large $L$ the corresponding potential energy has a linear confining behavior  $E \,=\, \sigma L \,$.
This result is reproduced in a cut-off $AdS$ space, considering a static string with end-points representing the  quark anti-quark pair. In this case  $\sigma$ is \cite{Boschi-Filho:2005mw}

\begin{equation}
\sigma = \frac{1}{2\pi\alpha' }  \,\frac{R^2}{z_0^2}\,\,.
\end{equation}

Using this result we express the free energy as

\begin{equation}
\label{FADS2}
F_{AdS} \, \sim \,-\,\sqrt{\sigma}\, \frac{ N^2 \sqrt{\alpha' } }{ R^4 }  \,\,.
\end{equation}

\noindent 

The string tension is related to the 't Hooft constant $ \,g\, N\, $. In the 3-d case where the string coupling constant $g$ has dimension of mass,  we have $\sqrt{\sigma} \sim g N$.
In the 4-d case, that we are considering  here, $g$ is dimensionless. From lattice studies one obtains $ l \sqrt{\sigma} \sim g(l) N  \,$ where  $l$ is a fixed length scale \cite{Lucini:2001ej}. 
So, for a fixed scale the effective string tension is independent of $N$ in the 
't Hooft limit ($gN$ fixed) and the free energy of the cut-off $AdS$ background is proportional to $N^2$. 
This result is similar to the case of $AdS$ in global  coordinates.

The expectation value of the energy is $ \langle E \rangle \,=\, - \frac{\partial }{ \partial \beta }
\,log Z \,\simeq \, \frac{\partial }{ \partial \beta } \,I_{total}\,\,$.
So we get 

\begin{eqnarray}
\langle E \rangle_{AdS} &=& F_{AdS}\,\,\,\,\,\sim N^2\,
\,\,\,\,\,\,\,\,\,\,\,\,(T < T_C )
\label{Energy1}\\
\nonumber\\
\langle E \rangle_{BH} &=&  
  \frac{3 R^3}{ 16  G_5 } \,  \pi^3 T^4  \,=\, \frac{3 N^2 \pi^2}{8  } \,  T^4 
 \,\,\,\,\,\,\,\,\,\,\,\,(T > T_C )  \,.
 \label{Energy2}
\end{eqnarray}

From these energies we calculate the entropies as $ S = \beta \langle E \rangle \, + log Z \,\simeq \,
\beta \langle E \rangle \, - I_{total}\,$ and find
\begin{eqnarray}
S_{AdS} &=& 0 \,\,\,\,\,\sim N^0\,\, \,\,\,\,\,\,\,\,\,\,\,\,(T < T_C)
\label{Entropy1}\\
\nonumber\\
S_{BH} &=& \, 
\frac{ R^3\,\pi^3 }{4  G_5 }\,T^3 \,=\,  \frac{ \pi^2 N^2  }{ 2  }\,T^3
 \,\,\,\,\,\,\,\,\,\,\,\,\,(T > T_C)\,.
\label{Entropy2}
\end{eqnarray}

\noindent So  we find the expected jump in the entropy representing the change 
of degrees of freedom in the confinement/deconfinement phase transition.

\section{ Soft Wall Model} 
Here we briefly discuss the phenomenologically motivated soft wall model \cite{Karch:2006pv} with the hope of 
showing that the entropy jump
does not  depend on the precise details of the model but rather on the general properties of the transition.
Other aspects of the thermodynamics of the soft wall model were discussed in \cite{Kajantie:2006hv}. 
This model consists in a background involving $AdS$ or black hole $AdS$ spaces and an effective dilaton field. The 
dilaton field does not backreact on the metric. This field essentially plays 
the role of a smooth infrared cut-off. 
The on-shell bulk action is

\begin{equation}
I_{bulk} \,=\,  \frac{4}{ R^2\,\kappa^2}\, \int_{{\cal M}} d^5 x
\sqrt{g} \,e^{-\Phi}\,\,, \label{Action4}
\end{equation}

\noindent where $\Phi \,=\, c\,z^2\,$ (the constant $c$ has dimension of mass squared). The bulk action
 densities for the $AdS$ and black hole are then

\begin{eqnarray}
I^{^{AdS}}_{bulk} &=& \frac{R^3 \beta }{\kappa^2}\,\Big[ \,
 c^2 ( \frac{3}{2} \,-\, \gamma ) \,+ \, \frac{1}{\epsilon^4 } 
- \frac{2 c}{\epsilon^2 } - c^2 \log ( c\epsilon^2 ) \,\,
\Big] \label{SoftVol1}
\\
\nonumber\\
I^{^{BH}}_{bulk} \,=\,  
\frac{R^3 \beta' }{\kappa^2}\,&\Big[& \,\,
 c^2 ( \frac{3}{2} \,-\, \gamma ) -c^2 Ei ( cz_h^2 ) 
\,+\, e^{-cz_h^2} \, \Big( \frac{c}{z_h^2} \,-\, \frac{1}{z_h^4} \,  \Big)
\nonumber\\
&+&  \frac{1}{{\epsilon'}^4 } 
- \frac{2 c}{{\epsilon'}^2 } - c^2 \log ( c{\epsilon'}^2 ) \,\, \Big]
 \label{SoftVol2}
\end{eqnarray}

\noindent where $\gamma \sim 0.5772 \,$ is the Euler-Mascheroni constant and $ Ei (u)\,\equiv \, \int_u^\infty \, exp(-t)/t \, dt \,$. Note that besides the $1/\epsilon^4\,$ pole, that appeared in the hard wall model, here we also have $1/\epsilon^2$ and $ \log ( c \, \epsilon^2 )$ poles.
These new singularities may also appear in the surface action and have to be canceled by the counterterms.
There is an ambiguity in the definition of surface terms and counterterms since the interaction between the dilaton field and the metric is not taken into account in this phenomenological model. 
We will follow a similar approach to that used for writing the bulk term.
We take the surface and counterterm actions of  section {\bf II} and multiply by a function of the dilaton field.  
The main fact is the existence of surface and counterterms such that 
their combination cancels the singularities of the bulk term. The dependence of 
the entropy on $N$ will not be affected by the precise form of these terms. The relevant terms that contribute to the surface and counterterms are the following

\begin{equation}
I_{surface}\,+\, I_{ct} \,=\, \, \frac{1}{\kappa^2\,R }\, \int_{{\cal \partial M}} d^4 x
\sqrt{h} \, \Big[ \,-1 \,+\,  2 \Phi \, + \, \log (\Phi ) \, \Phi^2 \,\Big] \label{SoftGeralCT}\,\,.
\end{equation}

This leads to  

\begin{eqnarray}
I^{^{AdS}}_{surface}\,+\, I^{^{AdS}}_{ct} &=& \frac{R^3 \beta }{\kappa^2}\,\Big[ \,
 - \, \frac{1}{\epsilon^4 } \,+ \, \frac{2 c}{\epsilon^2 }\,+\,  c^2 \log ( c\epsilon^2 ) \,\,
\Big], \label{SoftSurfCT}\\ \nonumber\\
I^{^{BH}}_{surface} \,+\, I^{^{BH}}_{ct}
&=&  \frac{R^3 \beta' }{\kappa^2}\,\Big[ \,\,
 \,-\,  \frac{1}{{\epsilon'}^4 } 
+\,  \frac{2 c}{{\epsilon'}^2 } +  c^2 \log ( c{\epsilon'}^2 ) \,
+ \frac{1}{2 z_h^4}
\, \Big]\,\,.
 \label{SoftSurfCT}
\end{eqnarray}

The total actions are

\begin{eqnarray}
I^{^{AdS}}_{total} &=& \frac{R^3 \beta }{\kappa^2}\,\Big[ \,
 c^2 ( \frac{3}{2} \,-\, \gamma ) \,\Big] \label{Softtotal1}
\\
\nonumber\\
I^{^{BH}}_{total} \,=\,  
\frac{R^3 \beta' }{\kappa^2}\,&\Big[& \,\,
 c^2 ( \frac{3}{2} \,-\, \gamma ) -c^2 Ei ( cz_h^2 ) 
\,+\, e^{-cz_h^2} \, \Big( \frac{c}{z_h^2} \,-\, \frac{1}{z_h^4} \,  \Big)
+  \frac{1}{2z_h^4}\,\, \Big]\,\,.
 \label{SoftTotal2}
\end{eqnarray}

The difference between these finite action gives the same result found in ref \cite{Herzog:2006ra}.
So there is a Hawking-Page transition corresponding to a confinement/deconfinement phase transition
as described there.
We could calculate all the thermodynamical variables from these actions as we did in the hard wall case.
But it is easy to see that the entropy has the expected jump. For the $AdS$ case (low temperature)
the entropy is zero, since the action is linear in $\beta$.
For the black hole case (high temperature), assuming that $c$ is $N$ independent, we find the same $N^2 $ 
dependence as  for the hard wall model.  Note that $ z_h =  \beta^\prime / \pi $ so the black hole action 
contains terms that are not linear in $\beta^\prime $. These are the terms that contribute to the black hole  entropy.

 Note that it is possible to introduce other non singular counterterms like $\,\sqrt{h}\,\Phi^2\,$ in
equation (\ref{SoftGeralCT}). However, these terms would not change the entropies since they are linear in $\beta$ or $\beta^\prime$. Also, they do not affect the Hawking Page transition because for $\beta = \beta' $
they are the same for both spaces.

\section{Conclusions}
In this paper we have revisited the thermodynamics of a very successful toy model in the context of the gauge/gravity 
correspondence. We have used holographic renormalization to compute the finite actions of the relevant 
supergravity backgrounds and verify the presence of a phase transition. We also computed the thermodynamical
variables for these solutions. We have shown that the entropy, in the gauge theory side, jumps from $N^0$ to $N^2$, a result that fits nicely with the expectation of  confinement/deconfinement phase transitions. 
We have also discussed the soft wall model and found the same entropy jump, 
indicating the universality of  our results. 

There are questions that we believe deserve further analysis. 
An important issue that we did not address in this note but that certainly deserves further attention is the following. 
The five dimensional gravity action (\ref{Action1}) is the reduction of ten-dimensional type IIB supergravity action. The main contribution that we included in (\ref{Action1}) is the cosmological constant $\Lambda$ which comes from the five-form field strength. However, 
it is known that in the Klebanov Strassler (KS)  and Maldacena Nunez (MN) models the  three-form field strengths play a crucial role. It would be interesting to clarify 
the role of these three-form field strengths from the five dimensional point of view. Finally, we would like to 
point out that the transition in the backgrounds that inspired the hard wall model can be addressed directly. Using the 
supergravity background constructed in \cite{PandoZayas:2006sa}, one would expect a similar Hawking-Page type transition between 
this background and the KS solution. There are other models of deconfinement and chiral symmetry restoration \cite{Aharony:2006da}
that deserve further study along these lines. 
Some of these issues are currently under investigation. 
 
\acknowledgements CABB,  HBF and NRFB are partially supported by CLAF, CNPq and FAPERJ. 
LPZ is  partially supported by Department of Energy under 
grant DE-FG02-95ER40899 to the University of Michigan and thanks UFRJ and CBPF for hospitality. 
We thank Michael Teper for important correspondence.

 \end{document}